\documentclass[12pt]{article}
\usepackage{putex}
\usepackage{graphicx}
\begin{document}

\preprint{hep-th/0611272 \\ PUPT-2217}

\institution{PU}{Joseph Henry Laboratories, Princeton University, Princeton, NJ 08544}

\title{Comparing the drag force on heavy quarks in ${\cal N}=4$ super-Yang-Mills theory and QCD}

\authors{Steven S.~Gubser}

\abstract{Computations of the drag force on a heavy quark moving through a thermal state of strongly coupled ${\cal N}=4$ super-Yang-Mills theory have appeared recently \cite{Herzog:2006gh,Casalderrey-Solana:2006rq,Gubser:2006bz}.  I compare the strength of this effect between ${\cal N}=4$ gauge theory and QCD, using the static force between external quarks to normalize the 't~Hooft coupling.  Comparing ${\cal N}=4$ and QCD at fixed energy density then leads to a relaxation time of roughly $2\,{\rm fm}/c$ for charm quarks moving through a quark-gluon plasma at $T=250\,{\rm MeV}$.  This estimate should be regarded as preliminary because of the difficulties of comparing two such different theories.}

\PACS{}
\date{November, 2006}

\maketitle

\section{Introduction}
\label{INTRODUCTION}

It was shown in \cite{Herzog:2006gh,Gubser:2006bz} that the drag force on a heavy quark moving through a thermal state of $SU(N)$ ${\cal N}=4$ super-Yang-Mills theory is
 \eqn{DragForce}{
  F_{\rm drag} = -{\pi \sqrt{g_{YM}^2 N} \over 2} T^2 
    {v \over \sqrt{1-v^2}} \,,
 }
in the limit of large $N$ and large $g_{YM}^2 N$.  Using $p/m = v/\sqrt{1-v^2}$ and $F_{\rm drag} = dp/dt$, one finds that momentum of the heavy quark falls by $1/e$ in a time
 \eqn{DragTime}{
  t_D = {1 \over \eta_D} = {2m \over \pi \sqrt{g_{YM}^2 N} T^2}
 }
where $\eta_D$ is the friction coefficient.\footnote{Subtleties regarding the dispersion relation for the heavy quark were considered in \cite{Herzog:2006gh}, but they do not affect \eno{DragTime} except to specify that $m$ is the ``kinetic mass.''}  In \cite{Casalderrey-Solana:2006rq}, a somewhat different approach to the non-relativistic limit was considered which allows the extraction of a diffusion coefficient for heavy quarks,
 \eqn{DiffusionCoef}{
  D = {2 \over \pi T \sqrt{g_{YM}^2 N}} \,.
 }
This result may also be recovered from \eno{DragForce} \cite{Herzog:2006gh,Casalderrey-Solana:2006rq}.  In a parallel development \cite{Liu:2006ug}, a partially light-like Wilson loop was calculated using an AdS/CFT prescription and related to the jet-quenching parameter for light quarks.\footnote{A significant literature has arisen extending the calculations of \cite{Liu:2006ug,Herzog:2006gh,Casalderrey-Solana:2006rq,Gubser:2006bz} and incorporating new insights and possible connections with the data from RHIC.  The recent contribution \cite{Chernicoff:2006yp} includes a better guide to this literature than I am able to provide here.}

The question naturally arises: What value of the 't~Hooft coupling, $g_{YM}^2 N$, should be used in comparing \eno{DragForce}, \eno{DragTime}, or \eno{DiffusionCoef} to data from RHIC?  ${\cal N}=4$ super-Yang-Mills theory is simply a different theory from QCD, so there may not be a clean answer.  Making comparisons at temperatures above the scale of confinement and chiral symmetry breaking in QCD at least avoids the most obvious roadblocks, namely that ${\cal N}=4$ doesn't confine and has no chiral symmetry breaking.  But in the regime of temperatures that RHIC can attain, it is understood from lattice simulations that there are significant departures from conformal invariance.  Also, ${\cal N}=4$ super-Yang-Mills has adjoint scalars which interact as strongly with external quarks as do the gauge bosons.

A natural starting point is to prescribe $g_{YM} = g_s$, where $g_{YM}$ is the coupling in ${\cal N}=4$ gauge theory and $g_s$ is the coupling in the QCD lagrangian, so that $\alpha_s = g_s^2/4\pi$.  The reason for this choice is that Feynman diagrams involving only gluons would then have the same amplitude---leaving aside issues of the renormalization group flow of the coupling.  However, it was pointed out in \cite{Caron-Huot:2006te} that at weak coupling, the prescription $g_{YM} = g_s$ results in a substantially larger Debye mass $m_D$ for ${\cal N}=4$ super-Yang-Mills than for QCD, and an even larger modification of a parameter $m_\infty$ entering into the finite-temperature dispersion relation for a highly relativistic particle.  Following different prescriptions, such as equating $m_D$ or $m_\infty$, means taking $g_{YM} = g_s/2$ or $g_s/3$.  But it was not claimed in \cite{Caron-Huot:2006te} that one can carry one of these last two relations directly to the strong coupling regime, nor indeed was it made clear how to implement the underlying ideas (matching of $m_D$ or $m_\infty$) in the strongly coupled regime.

The purpose of this note is to make two suggestions: first, normalize the 't~Hooft coupling by comparing the force between static quark and anti-quark to the predictions of more conventional methods, for example lattice gauge theory.  Second, compare QCD and ${\cal N}=4$ at fixed energy density rather than fixed temperature.  As we will see, these suggestions are not without their own problems, but the resulting comparison scheme has some physical motivation.  It results in somewhat lower estimates of the drag force than the naive prescription $g_{YM} = g_s$.\footnote{String theory estimates of related quantities have appeared, even previous to \cite{Herzog:2006gh,Casalderrey-Solana:2006rq,Gubser:2006bz}: In \cite{Sin:2004yx}, the opacity length of ${\cal N}=4$ to colored probes was argued to be $1/\pi T$, and estimated as $1/3\,{\rm fm}$ at RHIC.  In an approach complementary to the present paper, the same authors have argued that drag force is related to the magnetic string tension \cite{Sin:2006yz}.}

\section{Normalizing against the static Coulomb force}
\label{COULOMB}

The string theory computation of the force between quark and anti-quark is the well-known Wilson loop construction in the $AdS_5$-Schwarzschild geometry, first explored in \cite{Rey:1998ik,Maldacena:1998im} for the zero-temperature case, and in \cite{Rey:1998bq,Brandhuber:1998bs} for non-zero temperature (see also \cite{Witten:1998zw}).  The results are simple to state: the zero-temperature potential is
 \eqn{ZeroTV}{
  V(r) = -\sqrt{g_{YM}^2 N} {4\pi^2 \over \Gamma(1/4)^4} {1 \over r}
    \,,
 }
and for non-zero temperature the free energy and radius can be parametrically expressed as
 \eqn{QQbarParametric}{
  r &= {2 \over \pi T} \tilde{r}(q) \qquad
   F = T \sqrt{g_{YM}^2 N} \tilde{F}(q)  \cr
  \tilde{r}(q) &\equiv q \sqrt{1-q^4} \int_0^1 du \, 
    {u^2 \over \sqrt{1-u^4} \sqrt{1-q^4u^4}}  \cr
  \tilde{F}(q) &\equiv {1 \over q} \left[ 
    \int_0^1 {du \over u^2} \, \left( \sqrt{1-q^4u^4 \over 1-u^4} - 1
      \right) - 1 + q \right] \,,
 }
where for convenience we have introduced dimensionless forms $\tilde{r}$ and $\tilde{F}$ of the distance between the quarks and the free energy.  The dimensionless parameter $q$ is related to how far down into $AdS_5$-Schwarzschild the string dangles.  The integrals in \eno{QQbarParametric} can be done in terms of hypergeometric functions.  $F$ is negative only for $q < q_* \approx 0.66$, corresponding to $\tilde{r} < \tilde{r}_* \approx 0.38$.  In this range of (rescaled) radii, the preferred configuration is a single string joining the two external quarks.  For $\tilde{r}>\tilde{r}_*$, the preferred string configuration is two parallel (or, more properly, anti-parallel) strings stretching straight down into the horizon.  The conclusion, then, is that the quark-anti-quark force drops abruptly to zero for $\tilde{r}>\tilde{r}_*$.  This is unlike the exponential behavior characteristic of Debye screening.  The difference can probably be ascribed to the large $N$ limit: there really are attractive interactions between the anti-parallel strings that dominate the $\tilde{r}>\tilde{r}_*$ regime, but they are subleading in $N$ because they involve exchange of gravitons and other massless particles in $AdS_5$-Schwarzschild.

A representative temperature for the QGP created in a central gold-gold collision at RHIC is $T = 250\,{\rm MeV}$.  Keeping in mind that this is only a rough figure, and comparing QCD to ${\cal N}=4$ at the same temperature, one finds that a rescaled radius $\tilde{r}_*$ corresponds to $r_* = 0.19\,{\rm fm}$.

Lattice computations were done, for example, in \cite{Petreczky:2004pz} for QCD with three flavors, and in \cite{Kaczmarek:2005ui} for QCD with two flavors; see also \cite{Karsch:2006sf} for a review and \cite{Petrov:2006pf} for some very recent results.  Different measures of the screening length may be defined.  A Debye radius defined by examining the large $r$ behavior of the lattice free energy in the presence of quark and anti-quark is found in \cite{Kaczmarek:2005ui} to be roughly $r_D = 0.24\,{\rm fm}$ at $T = 250\,{\rm MeV}$ (taking $T_c = 190\,{\rm MeV}$).\footnote{The value of $T_c$ for QCD as determined by lattice simulations has some uncertainty.  I have chosen a value in line with the recent work \cite{Cheng:2006qk}.  However, in \cite{Aoki:2006br}, significantly lower values were obtained.  Revising $T_c$ down to $170\,{\rm MeV}$---which is within the range of values often quoted in recent years and closer to values found in \cite{Aoki:2006br}---would not greatly change the conclusions of this study.  For example, with $T_c = 170\,{\rm MeV}$ one would find $r_D = 0.21\,{\rm fm}$, $r_{\rm med} = 0.37\,{\rm fm}$, and $r_{\rm max}=0.28\,{\rm fm}$ at $T = 250\,{\rm MeV}$ from \cite{Kaczmarek:2005ui}.  These values, although lower than the ones quoted in \eno{SeveralR}, are not low enough to agree with string predictions for ${\cal N}=4$.  Fixed energy density comparisons, as implemented in \eno{CompareT}, thus can still be motivated by their tendency to partially ameliorate the discrepancy between screening lengths in ${\cal N}=4$ and QCD.}  This is pleasantly close to $r_*$ as computed in the previous paragraph, but perhaps a fairer point of comparison among the quantities considered in \cite{Kaczmarek:2005ui} is $r_{\rm med} \approx 0.41\,{\rm fm}$, defined as the radius where the zero-temperature potential equals the large-distance limit of the free energy.  Another scale of interest is $r_{\rm max} \approx 0.33\,{\rm fm}$, defined as the radius where a quantity $\alpha_{q\bar{q}}(r)$, to be discussed further below, is maximized.  (The values quoted were extracted from figures~8 and~11 of \cite{Kaczmarek:2005ui}).

It is not surprising that the screening length at fixed temperature should be smaller in ${\cal N}=4$ super-Yang-Mills than in QCD: just counting degrees of freedom gives about three times as many in the former than the latter.  To be more precise, if $g_*$ is defined through $\epsilon = g_* {\pi^2 \over 30} T^4$, then $g_* = 120$ in the weakly coupled limit of ${\cal N}=4$ with gauge group $SU(3)$, as compared to $g_* = 37$ for QCD with two flavors in the weak coupling limit and $g_* = 47.5$ for three flavors; and $g_* \approx 90$ in strongly coupled ${\cal N}=4$ (based on the $3/4$ factor of \cite{Gubser:1996de}), versus $g_* \approx 33$ for QCD with $2+1$ flavors (based on figure~14 of \cite{Karsch:2001cy}).  Because $\epsilon \propto T^4$ exactly in ${\cal N}=4$ (as a consequence of exact conformal invariance) and approximately in QCD for $T \gsim 1.1 T_c$, in order to have $\epsilon_{\rm SYM} = \epsilon_{\rm QCD}$ we should take
 \eqn{CompareT}{
  T_{\rm SYM} \approx 3^{-1/4} T_{\rm QCD} \,.
 }
For example, $T_{\rm QCD}=250\,{\rm MeV}$ corresponds to $T_{\rm SYM}=190\,{\rm MeV}$ according to \eno{CompareT}.

The choice \eno{CompareT} goes partway toward fixing the discrepancy in screening lengths: for $T_{\rm QCD}=250\,{\rm MeV}$ one finds
 \eqn{SeveralR}{
  r_* &= 0.25\,{\rm fm}  \cr
  r_D &\approx 0.24\,{\rm fm}  \cr
  r_{\rm med} &\approx 0.41\,{\rm fm}  \cr
  r_{\rm max} &\approx 0.33\,{\rm fm} \,.
 }
An alternative would be to choose $T_{\rm SYM}$ so that $r_*$ matches $r_{\rm med}$ or $r_{\rm max}$ exactly.  I am uncomfortable with this because the various quantities in \eno{SeveralR} are various ways of assigning a typical scale to functional forms that are significantly different between ${\cal N}=4$ and QCD.  Energy density is at least an unambiguous quantity for comparison.

Because $r_*$ has no dependence on $g_{YM}^2 N$ (at least in the leading large $N$, large $g_{YM}^2 N$ approximation), it gives no leverage for normalizing the 't~Hooft coupling.  To make progress one must compare the magnitude of the static quark-anti-quark force between string theory and the lattice.  The quantity I will focus on from lattice simulations is
 \eqn{alphaDef}{
  \alpha_{q\bar{q}}(r) = {3 \over 4} r^2 {dV \over dr} \qquad
   \hbox{or}\qquad
  \alpha_{q\bar{q}}(r,T) = {3 \over 4} r^2 
    {\partial F \over \partial r} \,,
 }
for zero and non-zero temperature, respectively.  ($F$ is the free energy.)  At zero temperature, the quark-anti-quark force is well approximated by Coulomb plus linear terms: for example, from \cite{Kaczmarek:2005ui},\footnote{By way of comparison, in \cite{Petreczky:2004pz} one finds
 \eqn[c]{OtherCornellPotential}{
  V(r) = -{4\alpha(r)/3 \over r} + \sigma r  \cr
   \alpha(r) = 0.33 - {0.003\,{\rm fm} \over r} \qquad
   \sigma = (470\,{\rm MeV})^2 = (0.42\,{\rm fm})^{-2}
 }
for radii $0.09\,{\rm fm} \leq r \leq 0.3\,{\rm fm}$.  This is for three flavors, wheras \eno{CornellPotential} is for two flavors.  Note also that the fitting region leading to \eno{OtherCornellPotential} is signicantly narrower from the one leading to \eno{CornellPotential}, and that \eno{OtherCornellPotential} was chosen rather than claimed to be a best fit form, though it appears in fact to be a good fit to lattice data over a slightly larger range than just described.}
 \eqn[c]{CornellPotential}{
  V(r) = -{4\alpha/3 \over r} + \sigma r  \cr
   \alpha = 0.212 \qquad
   \sigma = (420\,{\rm MeV})^2 = (0.47\,{\rm fm})^{-2}
 }
in the singlet channel for radii $0.1\,{\rm fm} \lsim r \lsim 1.2\,{\rm fm}$.  The origin of the definition \eno{alphaDef} is made clear by comparison to \eno{CornellPotential}: for radii on the small end of the specified window one has $\alpha_{q\bar{q}} \approx \alpha_s$.

\begin{figure}
  \centerline{\includegraphics[width=7.5in]{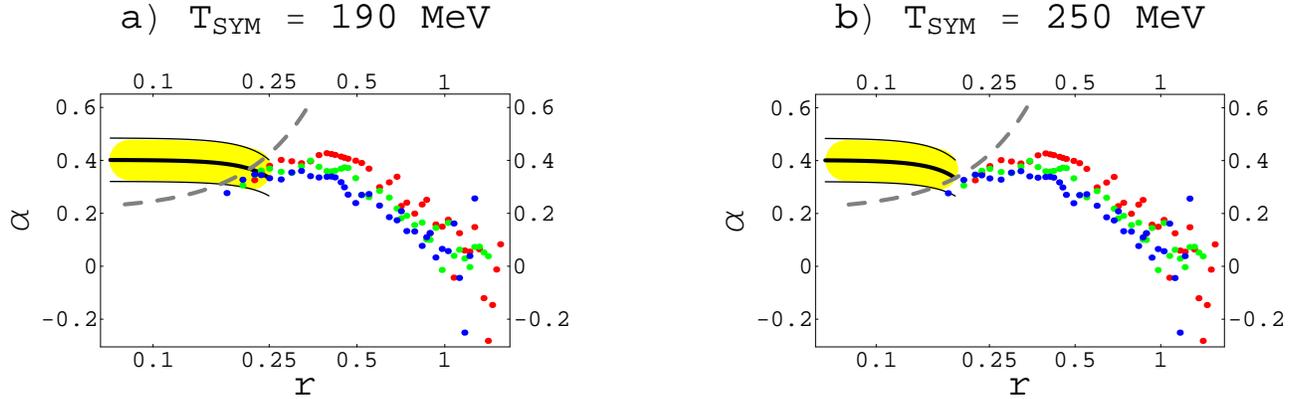}}
  \caption{Comparisons of the static force between quark and anti-quark between ${\cal N}=4$ super-Yang-Mills and QCD for two different values of $T_{\rm SYM}$.  The radius $r$, in ${\rm fm}$, is plotted on a log scale.  The thick black curve shows $\alpha_{\rm SYM}$ as defined by \eno{aYMdef} for $g_{YM}^2 N = 5.5$.  The thin upper curve is for $g_{YM}^2 N = 8$, and the thin lower curve is for $g_{YM}^2 N = 3.5$.  The dots are from lattice simulations \cite{Kaczmarek:2005ui} and are shown without error bars.  The red dots (highest on average) are for $T/T_c = 1.23$; the green dots are for $T/T_c = 1.37$; and the blue dots (lowest on average) are for $T/T_c = 1.5$.  The dashed grey curve shows the zero-temperature $\alpha_{q\bar{q}}(r)$ derived from \eno{CornellPotential}.}\label{CompareAlpha}
 \end{figure}

For non-zero temperature, $\alpha_{q\bar{q}}(r,T)$ must be evaluated from finite differences of evaluations of $F(r,T)$, so it is a quantity with some scatter.  In figure~\ref{CompareAlpha} a comparison is shown between $\alpha_{q\bar{q}}(r,T)$ from lattice simulations of two-flavor QCD \cite{Kaczmarek:2005ui}\footnote{I thank O.~Kaczmarek for providing me with the numerical values for the points shown in figure~\ref{CompareAlpha}.  Any errors in converting them to the format shown here are of course mine.} and the analogous quantity in ${\cal N}=4$ super-Yang-Mills,
 \eqn{aYMdef}{
  \alpha_{\rm SYM} = {3 \over 4} r^2 {\partial F \over \partial r} \,,
 }
with $F$ as defined in \eno{QQbarParametric}.
It is on the basis of the comparisons illustrated in this figure that I am going to use a range of values
 \eqn{tHooftRange}{
  3.5 = [g_{YM}^2 N]_{\rm lower} \lsim g_{YM}^2 N \lsim
    [g_{YM}^2 N]_{\rm upper} = 8 \,,
 }
and a representative value
 \eqn{tHooftTyp}{
  g_{YM}^2 N \sim [g_{YM}^2 N]_{\rm typ} = 5.5 \,.
 }
The value \eno{tHooftTyp} translates into a coupling $\alpha_{\rm SYM}^{\rm tree} \equiv g_{YM}^2/4\pi = 0.15$.  Note that this conventional definition of $\alpha^{\rm tree}_{\rm SYM}$ is rather different from the definition of $\alpha_{\rm SYM}$ in \eno{aYMdef}.

Several points deserve mention:
 \begin{enumerate}
  \item In construing the lattice results (which are for two-flavor QCD) as relevant to temperatures in the vicinity of $250\,{\rm MeV}$ I am assuming some degree of universality among lattice results for $\alpha_{q\bar{q}}$ as a function of $T/T_c$ for different numbers of flavors.  Such universality has been observed for related quantities in \cite{Kaczmarek:2005ui}.  Results for $\alpha_{q\bar{q}}(r,T)$ do not seem to be available for three-flavor QCD.  Scaling $T_c$ to $190\,{\rm MeV}$ means that the red, green, and blue dots correspond, respectively, to $234\,{\rm MeV}$, $260\,{\rm MeV}$, and $285\,{\rm MeV}$.\footnote{Choosing the lower value $T_c=170\,{\rm MeV}$ means that the red, green, and blue dots in figure~\ref{CompareAlpha} correspond, respectively, to $209\,{\rm MeV}$, $233\,{\rm MeV}$, and $255\,{\rm MeV}$.  The comparison with string theory is not greatly affected by this change: one should be matching the string theory predictions to the blue points rather than the red or green points.}\label{ScaleTc}
  \item In figure~\ref{CompareAlpha}a, I have followed \eno{CompareT} by choosing $T_{\rm SYM} = 190\,{\rm MeV}$ to compare with $T_{\rm QCD} = 250\,{\rm MeV}$.  In figure~\ref{CompareAlpha}b I reverted to the identification $T_{\rm SYM} = T_{\rm QCD}$ but used the same values of $g_{YM}^2 N$ as in figure~\ref{CompareAlpha}a; thus the string theory curves are simply translated to the left (note that radius is shown on a log scale).  It is amusing to note that by choosing a much lower value of temperature on the ${\cal N}=4$ side, for example $T_{\rm SYM} = 95\,{\rm MeV}$, a considerably better fit between $\alpha_{\rm SYM}$ and $\alpha_{q\bar{q}}(r,T)$ is obtained over a substantial range of scales, still with the same values of $g_{YM}^2 N$.\label{DifferentT}
  \item The super-Yang-Mills curve terminates at $r_* = 0.25\,{\rm fm}$ in figure~\ref{CompareAlpha}a and at $r_* = 0.19\,{\rm fm}$ in figure~\ref{CompareAlpha}b because for larger radii, the configuration that dominates the path integral is two disconnected strings whose attraction is subleading in the $1/N$ expansion, as discussed in the paragraph following \eno{QQbarParametric}.
  \item Because in figure~\ref{CompareAlpha}a there are only a handful of lattice points that overlap the range of the super-Yang-Mills curve, and in figure~\ref{CompareAlpha}b there are at most three, I have not performed a systematic fit to obtain \eno{tHooftRange} and \eno{tHooftTyp}, but rather chosen these values by eye to obtain an approximate match.
 \end{enumerate}

Evidently, the comparison of the static quark-anti-quark force between ${\cal N}=4$ and QCD leaves something to be desired.  Following up point~\ref{DifferentT} above, it is tempting to improve the ``fit'' by lowering $T_{\rm SYM}$ substantially.  But the optimal values for this purpose are roughly {\sl half} the value suggested in \eno{CompareT}, resulting in an energy density in ${\cal N}=4$ that is a factor of $16$ lower than in QCD.  Perhaps in light of the comparisons \eno{SeveralR} there is some case to be made for lowering $T_{\rm SYM}$ a bit below $T_{\rm QCD}/3^{1/4}$; but it doesn't make sense to me to throw thermodynamic comparisons so far out of whack as the choice $T_{\rm SYM} = T_{\rm QCD}/(2 \cdot 3^{1/4})$ would do.  What makes more sense is that, in order to get a better match between AdS/CFT predictions and lattice simulations, the former must include the effects of a running coupling.  The increase of $\alpha_{q\bar{q}}$ with $r$ that persists roughly up to $r_{\rm med} = 0.41\,{\rm fm}$ when $T/T_c = 1.37$ shows that the running coupling competes significantly against finite-temperature screening well above $r_D = 0.24\,{\rm fm}$.

To summarize: \eno{tHooftTyp} was obtained by comparing ${\cal N}=4$ and lattice QCD at fixed energy density by equating $\alpha_{\rm SYM}(r,T_{\rm SYM})$ and $\alpha_{q\bar{q}}(r,T_{\rm QCD})$ at roughly the largest $r$ for which $\alpha_{\rm SYM}(r,T_{\rm SYM})$ is defined.  It turns out that this radius is approximately $1/\pi T_{\rm QCD}$, using $T_{\rm QCD}=250\,{\rm MeV}$ (and keeping in mind the remarks about scaling $T_c$ as in point~\ref{ScaleTc} above).  This is not an unreasonable scale at which to try to make contact with the physics of charm quark diffusion in the QGP.  If ${\cal N}=4$ and QCD are compared at fixed temperature, there is less overlap between the range of $\alpha_{q\bar{q}}(r,T)$ and $\alpha_{\rm SYM}(r,T)$, but matching on the large $r$ end of that range yields values similar to \eno{tHooftTyp}.

With $N=3$ and $g_{YM}^2 N \approx 5.5$, both loop effects and stringy effects are likely to be significant, and fully including them could appreciably change the estimates \eno{tHooftRange} and~\eno{tHooftTyp}.  To gain a more quantitative appreciation of the size of stringy effects, note first that $g_{YM}^2 N = L^4/\alpha'^2$, so if \eno{tHooftTyp} is adopted, the curvature scale $L$ of $AdS_5$ is only a little larger than the string scale $\sqrt{\alpha'}$.  In \cite{Gubser:1998nz} it was argued that the free energy of the $AdS_5$-Schwarzschild background receives the following lowest-order stringy correction:\footnote{In \cite{Gubser:1998nz} as in other works of similar vintage, a different normalization convention was used for $g_{YM}$, such that $2 g_{YM}^2 N = L^4/\alpha'^2$.  I have duly adjusted this factor in expressing \eno{fCorrect} in the conventions of this paper.  Similar adjustments have been made in \eno{ZeroTV} and related formulas, and in \eno{etasValues}.}
 \eqn{fCorrect}{
  f(g_{YM}^2 N) \equiv {F \over F_{SB}} = {3 \over 4} + 
    {45 \over 32} \zeta(3) (g_{YM}^2 N)^{-3/2} \,,
 }
where $F_{SB}$ is the free energy of ${\cal N}=4$ super-Yang-Mills theory at zero coupling.  From \eno{tHooftTyp} and~\eno{fCorrect} one finds
 \eqn{fValues}{
  f(\left[ g_{YM}^2 N \right]_{\rm typ}) = 0.88 \,.
 }
So the corrections to purely thermodynamic quantities are fairly modest.

In \cite{Buchel:2004di} it was argued that the ratio of the shear viscosity to the entropy density has a stronger dependence on the 't~Hooft coupling:
 \eqn{etasValues}{
  {\eta \over s} = {1 \over 4\pi} \left( 1 + 
    {135 \over 8} \zeta(3) (g_{YM}^2 N)^{-3/2} \right) \,,
 }
including the first $\alpha'$ correction.  Plugging \eno{tHooftTyp} into \eno{etasValues} results in $\eta/s = 0.2$.  This may still be in an acceptable range for hydrodynamical models of elliptic flow as measured at RHIC.\footnote{I thank U.~Wiedemann for correcting a numerical error in the original version of this manuscript, and for remarks regarding the range of $\eta/s$ that is consistent with data.}  But because the correction term in \eno{etasValues} is larger than the leading term for $g_{YM}^2 N = 5.5$, higher order corrections must be expected to be significant.\footnote{For $g_{YM}^2 N = 6\pi$---corresponding to $\alpha_s=0.5$ in the ``obvious'' matching prescription $g_{YM} = g_s$, $N=3$, $T_{\rm SYM}=T_{\rm QCD}$---one finds $\eta/s \approx 1.2/4\pi \approx 0.10$ from \eno{etasValues}, i.e.~the correction from the $\zeta(3)$ term is less than a quarter the size of the leading effect.}

\section{Discussion}

As a baseline for comparisons of ${\cal N}=4$ and QCD, let us use the ``obvious'' prescription $g_{YM} = g_s$, $N=3$, $T_{\rm SYM} = T_{\rm QCD}$, and the often-quoted value $\alpha_s = g_s^2/4\pi = 0.5$, so that $g_{YM}^2 N = 6\pi$.  We will also use, throughout, the value $m=1.4\,{\rm GeV}$ for charm quarks and a temperature $T_{\rm QCD}=250\,{\rm MeV}$.  Then, from \eno{DragTime} and \eno{DiffusionCoef},
 \eqn{Baseline}{
  t_D \approx 0.6\,{\rm fm} \qquad 2\pi T D \approx 0.9 \,.
 }
If one uses instead the prescription of equal energy densities and $\alpha_{\rm SYM} = \alpha_{q\bar{q}}$ as implemented in~\eno{CompareT} and~\eno{tHooftTyp}, one obtains
 \eqn{NewValue}{
  t_D = {2m \over \pi \sqrt{[g_{YM}^2 N]_{\rm typ}} T_{\rm SYM}^2}
    \approx 2.1\,{\rm fm} \,.
 }
If, in QCD, one wishes to reproduce the damping time \eno{NewValue} using Langevin dynamics at a temperature $T=250\,{\rm MeV}$, then the value one needs is
 \eqn{FancyValue}{ 
  2\pi T D = {2\pi T_{\rm QCD}^2 \over m} t_D \approx 3.0 \,.
 }
A tricky point is that if one were to apply fluctuation-dissipation logic directly in ${\cal N}=4$ super-Yang-Mills, then instead of $T_{\rm QCD}$ in \eno{FancyValue} one would naturally use $T_{\rm SYM}$.  The result for $2\pi T D$ would then be $1.7$.  This underscores a disadvantage of comparing at fixed energy rather than fixed temperature: factors of temperature are (for good reason) more often used to set scales and form dimensionless ratios.  Nevertheless, comparing the damping time $t_D$ is a defensible choice because it directly captures the drag on heavy quarks from the medium.  Therefore I stick with \eno{NewValue} and \eno{FancyValue} as representative of the fixed energy density comparison.

If a fixed temperature comparison is preferred, then using again the value quoted in \eno{tHooftTyp} for the 't~Hooft coupling leads to
 \eqn{FixedTempComparison}{
  t_D = 1.2\,{\rm fm}/c \qquad
   2\pi T D = 1.7 \,.
 }

The factor of $3.2$ between \eno{Baseline} on one hand and \eno{NewValue} and \eno{FancyValue} on the other comes from two effects: comparing at fixed energy density introduces a factor of $\sqrt{3}$, and a slightly larger factor comes from changing $g_{YM}^2 N$ from $6\pi$ to $5.5$.  The procedure of normalizing the 't~Hooft coupling by comparing the static force between quark and anti-quark seems to me quite well motivated.  The fixed energy density comparison is less so.  Recall that it was introduced to ameliorate a disparity in screening scales between string and lattice calculations, but its leverage on that front is only a factor of $3^{1/4} \approx 1.3$, whereas its final impact on the damping time $t_D$ is a factor of $\sqrt{3} \approx 1.7$.  Nevertheless, {\sl some} factor significantly greater than unity in $t_D$ is appropriate to reflect the smaller field content and charge assignments of QCD as compared to ${\cal N}=4$.  Altogether, the theoretical uncertainties of comparing ${\cal N}=4$ to QCD should be understood as implying a substantial uncertainty in the result \eno{NewValue}: briefly, $t_D = 2.1 \pm 1 \, {\rm fm}/c$ for $T_{\rm QCD} = 250\,{\rm MeV}$.

One may ask whether the value \eno{tHooftTyp} isn't an underestimate: if the string theory curves in figure~\ref{CompareAlpha} are continued ``by eye'' to larger radii, it would seem that a larger $g_{YM}^2 N$ would help the match.  I contend that such a continuation is entirely unsystematic, and I return to the point that a comparison between QCD and a string theory construction that incorporates a running coupling is called for.  If lattice results for static quark-anti-quark forces can be better described in such an approach, perhaps predictions for the drag on heavy quarks could be made with less uncertainty.

Readers familiar with the Wilson loop literature in AdS/CFT may be concerned that the results \eno{QQbarParametric} reflect not only gluon exchange between the quark and anti-quark, but also massless adjoint scalar exchange (and also the interactions of these fields with the rest of strongly coupled ${\cal N}=4$ SYM).  Should the free energy in \eno{QQbarParametric} be cut in half to reflect only the contribution of the gauge bosons?  I think not, because the drag force calculations in \cite{Herzog:2006gh,Gubser:2006bz}, just like the Wilson loop calculations that led to \eno{QQbarParametric}, apply to strings at a definite location in $S^5$.  So scalar couplings to the heavy quark contribute to the drag force.\footnote{It is tempting to think that one may ``average'' over the position of the string on $S^5$ in such a way as to eliminate the scalar charge and be left with only gauge interactions with the heavy quark.  But I don't think there is a well-understood way to do this.  One way of saying it is that the obvious averaging procedure amounts to computing the drag force for any given position on $S^5$ and then averaging the result afterwards, which doesn't change the scalar contribution.}

The experimental situation on charm's interaction with the QGP is not entirely clear.  The main experimental probe is the detection of energetic electrons and positrons coming from decays like $c \to s\bar{e}\nu$.  Some high $p_T$ electrons probably also come from $b$ decays.  Recent experimental accounts have appeared in \cite{Abelev:2006db,Adare:2006nq}, and earlier work includes \cite{Adler:2005xv,Bielcik:2005wu}.

Measurements of $R_{AA}$ and of $v_2$ for heavy quarks as reported in the most recent study from the PHENIX collaboration \cite{Adare:2006nq} do not appear to be consistent with Langevin simulations as performed in \cite{Moore:2004tg} for a single value of $D$.  The best agreement reported in \cite{Adare:2006nq} is with a model \cite{vanHees:2005wb} whose central assumption is the existence of D-like and B-like resonances in the QGP.  Interactions with these resonances dominate the energy loss of heavy quarks and were argued in \cite{vanHees:2004gq} to lead to relaxation times about three times smaller than perturbative QCD scattering processes predict.  If figure~4 of \cite{vanHees:2004gq} may be taken as representative of relaxation times in this approach, then at $T = 250\,{\rm MeV}$ the relaxation time is roughly $4.5\,{\rm fm}/c$, which is still somewhat larger than \eno{NewValue}.

Measurements of $R_{AA}$ for heavy quarks as reported in the most recent study from the STAR collaboration \cite{Abelev:2006db} show a somewhat swifter drop toward values around $0.2$ than the PHENIX study \cite{Adare:2006nq}.  $R_{AA} \sim 0.2$ is characteristic of measurements for charged hadrons, presumably dominated by events where the hard parton is a light quark or a gluon.  Agreement between \cite{Abelev:2006db} and \cite{vanHees:2005wb} is less good than between \cite{Adare:2006nq} and \cite{vanHees:2005wb}: the data in \cite{Abelev:2006db} seem to show swifter energy loss.  The best agreement reported in \cite{Abelev:2006db} is with calculations along the lines of \cite{Armesto:2005mz}, but for charm quarks only.  These calculations use a jet-quenching parameter $\hat{q} \approx 14\,{\rm GeV}^2/{\rm fm}$, which is in the upper range of expectations based on light quark data and difficult to understand perturbatively.\footnote{This value of $\hat{q}$ is also higher by about a factor of $3$ than the value $\hat{q}\approx 5\,{\rm GeV}^2/{\rm fm}$ obtained in \cite{Liu:2006ug} from AdS/CFT using a temperature of $310\,{\rm MeV}$.}  Altogether, I would interpret the recent comparisons of data to theoretical models as favoring charm quark relaxation times that are larger than \eno{NewValue} at $T_{\rm QCD}=250\,{\rm MeV}$, but smaller than one would expect based on any first-principles QCD calculation I am aware of.

\section{Conclusions}

The string theory prediction \cite{Herzog:2006gh,Casalderrey-Solana:2006rq,Gubser:2006bz} for the relaxation time of a charm quark propagating through a QGP with $T=250\,{\rm MeV}$ has been quoted as $0.6\,{\rm fm}/c$, but this value is based on a comparison prescription between ${\cal N}=4$ super-Yang-Mills and QCD which is not necessarily the best motivated physically.  The alternative prescription proposed herein comprises two aspects:
 \begin{itemize}
  \item Instead of comparing at fixed temperature, I have suggested comparing at fixed energy density.  The screening length in ${\cal N}=4$ is then closer to those found in lattice simulations.  Fixed temperature comparisons must be expected to lead to over-estimates in both screening and drag effects in QCD, due to the larger field content and charge assignments in ${\cal N}=4$.  Fixed energy comparisons provide a way---admittedly somewhat ad hoc---of partially correcting for this.
  \item Instead of comparing with $g_{YM} = g_s$, I advocate normalizing the 't~Hooft coupling of ${\cal N}=4$ by comparing the static force between quark and anti-quark between string theory and lattice simulations of QCD.  This makes sense because the string theory picture of the drag force is that a string trails out behind the quark which, if the quark had been static and an anti-quark had been nearby, would have mediated the static attraction between the two.
 \end{itemize}

With the comparison prescription just described, the string theory prediction for the relaxation time of a charm quark propagating through a QGP with $T_{\rm QCD}=250\,{\rm MeV}$ is $t_D = 2.1\,{\rm fm}/c$.  This value scales roughly as the inverse square of the temperature.  The comparison prescription, although physically motivated, cannot be regarded as a precise map, particularly in light of the departures from conformal invariance that QCD exhibits in the relevant temperature range.  I regard the estimate $t_D = 2.1\,{\rm fm}/c$ as uncertain by about $1\,{\rm fm}/c$ in either direction.

It is interesting to note that the thermalization time as estimated from quasinormal frequencies of the global $AdS_5$-Schwarzschild black hole in \cite{Friess:2006kw} rises from $0.3\,{\rm fm}/c$ to $0.4\,{\rm fm}/c$ if one employs a fixed energy density comparison as implemented through \eno{CompareT}.  These values are intriguingly close to the range $0.6-1.0\,{\rm fm}/c$ favored in hydrodynamic treatments of elliptic flow, but like \eno{NewValue} should be understood as incorporating considerable theoretical uncertainties.

\section*{Acknowledgments}

I thank G.~Michalogiorgakis and K.~Rajagopal for helpful discussions, and Z.~Fodor, O.~Kaczmarek, U.~Wiedemann, and B.~Zajc for useful e-mail exchanges.  I especially thank O.~Kaczmarek for providing me with numerical results from lattice simulations.  This work was supported in part by the Department of Energy under Grant No.\ DE-FG02-91ER40671, and by the Sloan Foundation.

\bibliographystyle{ssg}
\bibliography{strong}

\end{document}